\documentclass[
 reprint,amsmath,amssymb,aps,prl,
 superscriptaddress,
]{revtex4-2}

\usepackage{graphicx}
\usepackage{dcolumn}
\usepackage{bm}
\usepackage{hyperref}
\usepackage{color}

\begin{document}

\preprint{APS/123-QED}

\title{Quantum disorder induced by nuclear tunneling in lattice}

\author{Yu-Cheng Zhu}
\author{Jia-Xi Zeng}
 \email{hank@pku.edu.cn}
 \affiliation{State Key Laboratory for Artificial Microstructure and Mesoscopic Physics, Frontier Science Center for Nano-optoelectronics and School of Physics, Peking University, Beijing 100871, People's Republic of China}
 
\author{Qi-Jun Ye}
\affiliation{State Key Laboratory for Artificial Microstructure and Mesoscopic Physics, Frontier Science Center for Nano-optoelectronics and School of Physics, Peking University, Beijing 100871, People's Republic of China}
\affiliation{Interdisciplinary Institute of Light-Element Quantum Materials, Research Center for Light-Element Advanced Materials, and Collaborative Innovation Center of Quantum Materials, Peking University, Beijing 100871, People's Republic of China}

\author{Xin-Zheng Li}
 \email{xzli@pku.edu.cn}
\affiliation{State Key Laboratory for Artificial Microstructure and Mesoscopic Physics, Frontier Science Center for Nano-optoelectronics and School of Physics, Peking University, Beijing 100871, People's Republic of China}
\affiliation{Interdisciplinary Institute of Light-Element Quantum Materials, Research Center for Light-Element Advanced Materials, and Collaborative Innovation Center of Quantum Materials, Peking University, Beijing 100871, People's Republic of China}
\affiliation{Peking University Yangtze Delta Institute of Optoelectronics, Nantong, Jiangsu 226010, People’s Republic of China}
\date{\today}

\begin{abstract}
Lattice degrees of freedom (DoFs) may induce quantum disorder (QD) when nuclear tunneling outvies long-range order, but conventional phonon 
theory is incapable of describing such QD phases. 
Here we develop a method based on path-integral molecular dynamics to solve this problem. 
Its accuracy is verified in a double-well chain model and it is applied to a real material from first principles. 
A quantum order-disorder-order phase transition sequence is demonstrated when varying the strength of quantum fluctuations using the lattice constants as the tuning factor. 
Combining the excitation spectra and R\'enyi entanglement entropy, we pinpoint the QD region. 
This picture may be general in lattice systems having soft phonon modes, not limited to quantum paraelectricity, in which novel entangled lattice motion 
and its coupling with other DoFs can be expected. 

\end{abstract}

\maketitle

\begin{figure}[b]
	\includegraphics[width=8.5cm]{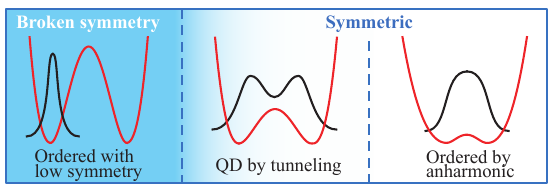}
	\caption{ 
		Schematic diagram of the order-disorder-order phase transition sequence. 
		The red and black curves sketch PES of soft modes and nuclear density distribution respectively. 
		When the barrier is high (low), the system has low (high) symmetry. 
		In between, there is a QD region. 
	}
	\label{Fig1}
\end{figure}

%
Quantum disorder (QD) is at the forefront of questing many-body physics and exploring novel quantum materials. 
Starting from the early days of quantum mechanics, QD has been proposed in spin models whose long-range order of ground state is suppressed by quantum fluctuations\cite{Kubo1953,PFEUTY1970,Amico2008,Sachdev2011}.
Quantum spin liquid, a prototypical example, has been a focus of research interest in recent years due to its potential in exhibiting exotic 
properties such as topological order and fractionalized excitation\cite{Savary2017,Zhou2017,Broholm2020}.
Apart from manifesting an intrinsic spin liquid in magnetic systems, budding attention is also paid to the lattice degrees of freedom (DoFs)\cite{shen2016,wang2020}. 
Existing experiments indicate that quantum paraelectricity (QPE) are QD phases\cite{Muller1979,muller1991}.
However, such cognition stays at a hypothetical level due to lack of description of quantum entangled lattice dynamics and clarification on QD region. 
One main difficulty for the absence of these studies is that conventionally the lattice DoFs are treated by phonon, which is perturbative and may fail in systems with soft phonon modes at low temperatures. 
The soft phonon mode indicates multiple degenerate wells on local Born-Oppenheimer potential energy surface (PES) and a structural instability\cite{Scott1974} as sketched in Fig.~\ref{Fig1}. 
When the barrier between the wells is not uncrossable, one can expect coupling between the wells which gives rise to uncertainty and quantum fluctuations (QF)\cite{BROUT1966507}. 
This coupling is characterized by nuclear tunneling\cite{Blinc1960}. 
Phonon is inefficient to expand the coupling and describe the quantum entanglement in this scenario. 
In fact, QD on the soft mode is speculated by a phenomenological analogy to models like quantum Ising chain\cite{DEGENNES1963132,PFEUTY1970,Tosatti1994,Zhong1996}. 
Missing of atomic details leads to controversies, e.g. some explanations of the soft modes are based on rattling whilst others are based on anharmonic correction\cite{Saikat2018,Tadano2018,Klarbring2020}. 
Evidences other than dielectric constants are required both experimentally and theoretically to clarify such controversies, especially for soft mode without an electrical dipole. 
In QPEs like $\textup{SrTiO}_3$, $\textup{KTaO}_3$ and $\textup{BaFe}_{12}\textup{O}_{19}$, it is already known that rich physics may arise due to competition between the nuclear QF and intersite interactions, which is sensitive to external factors, such as pressure, strain and doping\cite{sirenko2000,LinXiao2019,Maria2020,Zhang2020,Ranalli2023,Guo2012}. 
To clarify the rich physics behind such substance, a theoretical method beyond the conventional phonon theory for lattice dynamics is clearly needed. 

\begin{figure}
	\includegraphics[width=8.5cm]{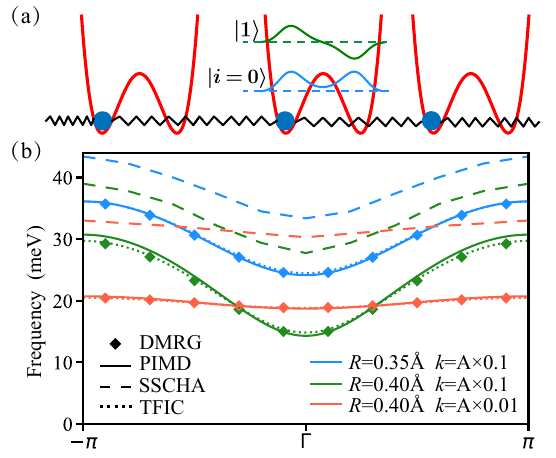}
	\caption{ 
		(a) A schematic diagram of 1-D double-well chain.
		One degree of freedom (the ball) lies on a double-well potential (red curve) connected to neighboring site by spring interaction (zigzag lines). 
		(b) Excitation spectra of 1-D double-well chain. The DMRG (diamonds), PIMD (solid lines) and TFIC (dotted lines) are in good agreement. SSCHA (dashed lines) overestimates the excitation energies because of its shortcoming in describing tunneling. 
		The blue (red) lines are spectra of chain with low (high) barrier and strong (weak) spring. 
	}
	\label{Fig2}
\end{figure}

Here, we develop a method for lattice dynamics to incorporate quantum entanglement from nuclear QF. 
This method is based on path-integral molecular dynamics (PIMD) which is mature for distinguishable continuous-space systems\cite{Cazorla2017,chandler1981,marx1996}. 
We obtain the excitation spectra using force constants on the free energy surface (FES) constructed by centroid potential of mean force instead of the PES. 
Its accuracy in describing QD phases is demonstrated in a double-well chain model by comparing with the exact solution. 
We then apply it to a real material with the Born-Oppenheimer PES constructed by combining $ab\ initio$ calculation and machine learning techniques. 
We find an order-disorder-order phase transition sequence with decreasing barrier for quantum tunneling.
From an ordered phase with low symmetry, the onset of QF drives the system into a QD phase (Fig.~\ref{Fig1}). 
Further reducing the barrier, the QF from nuclear tunneling vanishes and the system goes into an ordered phase with high symmetry. 
The first transition distinguishes between equilibrium structure with broken and high symmetry, and the second divides tunneling- and anharmonic-dominated phases. 
The two quantum phase transitions can be characterized by divergences of R\'enyi entanglement entropy. 
We start from a double-well chain model. 
This is a minimal model for order-disorder phase transitions, which has been used to study ferroelectrics, with one degree of freedom lying on a local double-well 
potential on each site and spring-like interaction between sites\cite{salje1991} (Fig.~\ref{Fig2}(a)).
In general, with a quartic double-well and quadratic spring, the Hamiltonian reads,
\begin{eqnarray}\label{H1}
	H = \sum_{n}  \left[ \frac{P_n^2}{2m} +  A (x_n^2-R^2)^2 +  \frac{1}{2} k  (x_n-x_{n+1})^2 \right]
\end{eqnarray}
where  $x_n$ is the relative position on site $n$, i.e. $x_n=u_n-nl$ with $l$ being the lattice constant. 
Each double-well has two local minima at $nl\pm R$. 
We set $A=0.00461~\textup{a.u.}$ and $m$ as the mass of a proton. 
$R$ and $k$ control the shape of barrier and the spring strength respectively. 
To simplify, we assume a periodic boundary and rewrite the Hamiltonian as
\begin{eqnarray}\label{H}
	H = \sum_{n}  \left[ \frac{P_n^2}{2m} +  A (x_n^2-\tilde{R}^2)^2 +   k  x_n x_{n+1} \right]
\end{eqnarray}
with a constant neglected and $\tilde{R}^2=R^2-k/2A$.
The local basis was chosen as the eigenstates $\left\lbrace  | i_n \rangle \right\rbrace$ of the reshaped double-well $h_n=P_n^2/2m +  A (x_n^2-\tilde{R}^2)^2$. 
The two lowest levels $i=0, 1$ are the symmetric and anti-symmetric superposition states respectively whose energy difference $\Delta=E_1-E_0$ is called the tunneling splitting.
Then the above Hamiltonian transforms to
\begin{eqnarray}\label{H2}
	\hat{H} = \sum_{n} \hat{h}_n
	-k \sum_{n} \hat{X}_n \ \hat{X}_{n+1}
\end{eqnarray}
with $\hat{h}_n = \sum E_{i} |i\rangle \langle i|$ and $\hat{X}_n = \sum_{i,j} | i \rangle \langle i | \hat{ x}_n | j \rangle \langle j |$.
The first onsite term provides QF by nuclear tunneling, while the second intersite term prefers ordering such as ferroelectricity. 
If retaining only the two lowest levels in local basis, one arrives at the transverse-field Ising chain (TFIC) $\hat{H}=\tilde{\Delta}\sum \hat{\sigma}_n^z-\tilde{k}\sum \hat{\sigma}_n^x \hat{\sigma}_{n+1}^x$, which gives a QD phase when $\tilde{\Delta}>\tilde{k}$~\cite{Sachdev2011,Glen2024}. 

Eq.~\ref{H2} can be precisely solved by the Density Matrix Renormalization Group (DMRG) method\cite{White1992,RMP2005,zhai_block2_2023}.
The first excitation band is shown in Fig.~\ref{Fig2}(b) for different parameters used in Eq.~\ref{H}.
The single-particle excitation bands of TFIC are very close to the DMRG ones, supporting that the ``two-levels'' approximation is reasonable in describing the disordered phase. 
Then we look at the solid lines from our new method based on PIMD. 
Its idea is that one can always extract frequency of a regular motion from the FES, with contributions from thermal and quantum fluctuations taken into account. 
At low temperatures, one just need force constants by several finite displacements around equilibrium structure to obtain excitation spectra. 
We get these force constants by centroid potential of mean force in a fixed-centroid PIMD, based on an efficient and high-quality sampling. 
As shown in Fig.~\ref{Fig2}(b), for a wide range of barrier height $V_{\textup{barrier}}=AR^4$ and spring strength $k$ in disordered phase, the PIMD lines are also in good agreement with the DMRG ones. 
With $V_{\textup{barrier}}$ increasing, the energy gap at $\Gamma$ point tends to close until a quantum order-disorder phase transition. 
As an example of neglecting tunneling effect, the stochastic self-consistent harmonic approximation (SSCHA)\cite{monacelli2021} is shown to overestimate the excitation energy, and 
the departure becomes serious near the critical point. 
Considering the fact that SSCHA is prevalent in calculating phonon spectra, we note that its inherent Gaussian density distribution of the nuclear wave function which precludes it from 
accurate description of tunneling\cite{Siciliano2024} means that one should be cautious for its application in such a scenario. 
Compared with DMRG and TFIC, one advantage of our PIMD method is that it is applicable to real materials.
So we apply it to $\textup{SrTiO}_3$ from first principles, with the Born-Oppenheimer PES constructed using a combination of density-functional theory (DFT) and machine 
learning force field (MLFF).
The DFT calculations are performed by the Vienna Ab initio Simulation Package (VASP)\cite{KRESSE1996}  
at GGA level using the Perdew-Berke-Ernzerhof (PBE) functional\cite{Perdew1996} 
and projector augmented wave (PAW) pseudopotentials\cite{Kresse1999}. 
The machine learning techniques are used to save computational cost so that our PIMD based phonon spectra calculation is doable\cite{Zhu2022,Loose2022,Fang2024}.
Here we employ the MLFF method built in VASP in view of its reliability for crystals as demonstrated by Verdi el al.\cite{verdi2023,Ranalli2023} 
As a practical test on its accuray, the harmonic phonon spectra (by Phonopy software\cite{phonopy2023}) in comparison with the DFT one is shown in Fig.~S1\footnote{\label{SM}See Supplemental Material at [URL] for more computational details and convergence tests.}.
Then, the PIMD simulations are performed using the i-pi software\cite{KAPIL2019} with self-developed implementations for excitation spectra and R\'enyi entropy calculations. 
We fix the centroid at finite-displaced structures (0.03 $\textup{\AA}$ from equilibrium position) when sampling centroid potential of mean force 
and define particular normal modes according to the ``replica exchange'' technique\cite{Hastings2010} when calculating the R\'enyi entropy. 
More computational details are shown in the Supplemental Material.\footnotemark[1]

\begin{figure}[b]
	\includegraphics[width=8.7cm]{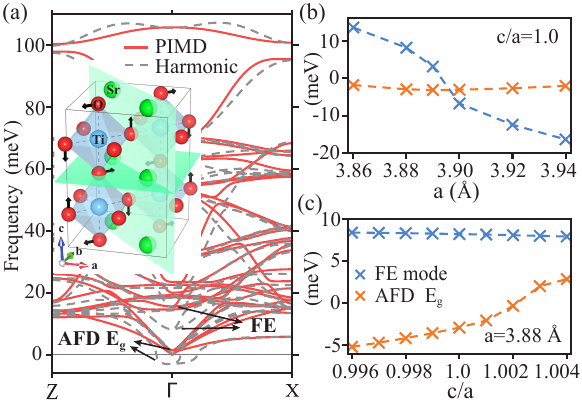}
	\caption{ 
		(a) The excitation spectra of tetragonal $\textup{SrTiO}_3$ with $a,c=3.88~\textup{\AA}$ by PIMD at 6~K (red lines) and harmonic phonon (gray dashed lines). 
		The AFD mode is lifted from an imaginary frequency to $\sim$2 meV. 
		The inset shows a unit cell, in which the black arrows sketch the $E_g$ component of AFD mode. 
		(b) and (c) show Frequencies of AFD and FE modes by harmonic phonon. They are sensitive to $c/a$ and $a$ respectively. 
	}
	\label{Fig3}
\end{figure}

$\textup{SrTiO}_3$ has a cubic perovskite phase ($Pm\bar{3}m$) at room temperature, which transforms to a tetragonal one ($I4/mcm$) below 105 K. 
This first-order phase transition is driven by the $A_{1g}$ component of an anti-ferrodistortive (AFD) mode where two adjacent oxygen octahedras in $ab$ plane rotate reversely around $c$ axis\cite{Fleury1968}. 
From DFT, the $E_g$ components of AFD mode (as marked by black arrows in inset of Fig.~\ref{Fig3}(a)) is soft at $T=0~\textup{K}$, similar to the ferroelectrical (FE) mode. 
This soft FE mode is responsible for its QPE\cite{Zhong1996}. 
The AFD $E_g$ mode (referred as AFD mode below) is sensitive to the uniaxial tension of lattice, whilst the FE mode is coupled with uniform pressure (Fig.~\ref{Fig3}(b, c))\cite{Zhong1995,Shin2021}.
Thus, a tiny change in lattice constants would affect the softness of these modes and $\textup{SrTiO}_3$ is an incipient ferroelectric. 
The strain sensitivity also means that it is of good operability to control the softness of these modes and related properties. 
For a clear demonstration of the picture of QD, other than the FE mode, we focus on the AFD mode which is more representative of general soft modes in lattice without electric dipoles.  
This is enabled by increasing pressure to exclude the softness of FE mode, 
so that only the AFD mode is soft. 
%
%
As in Fig.~\ref{Fig3}(a), with the lattice constants $a=3.88~\text{\AA}$ and $c/a=1.0$, the AFD (FE) mode has an imaginary (real) frequency in harmonic phonon spectrum. 
The spectra of primitive cell are shown by the gray dashed (red solid) lines for the harmonic (PIMD) results. 
%
%
Except for the AFD and FE modes, other optical phonon ones are just slightly corrected by PIMD. 
Although without a local degeneracy, the FE mode is still lifted to a higher frequency because of strong scattering with other phonon. 
This anharmonic effect can be described by methods with atomic details like SSCHA but is neglected in model studies like TFIC or Vanderbilt model Hamiltonian\cite{Zhong1996}. 
However, the SSCHA has been shown to fail for tunneling effect. 
So, a unified theory from first principles is indispensable to quantify the delicate interplay between tunneling and anharmonic effects. 
The soft AFD mode is also lifted to a positive frequency with a gap at $\Gamma$ point of about $2~\textup{meV}$. 
It is unclear whether this lift comes from tunneling or anharmonic effect, which leads to a controversy.  
To clarify this, we vary the softness of AFD mode by $c/a$ to control the strength of nuclear QF. 
Meanwhile, the frequencies of modes other than AFD remain unchanged. 
So we can separate variables and focus on AFD mode in the following. 
We perform PIMD simulations using a $3\times3\times3$ supercell of unit cell to mitigate the finite size effect. 
The unit cell of tetragonal $\textup{SrTiO}_3$ contains 20 atoms and has lattice constants of $(\sqrt{2}a, \sqrt{2}a, 2c)$. 

\begin{figure}
	\includegraphics[width=8.5cm]{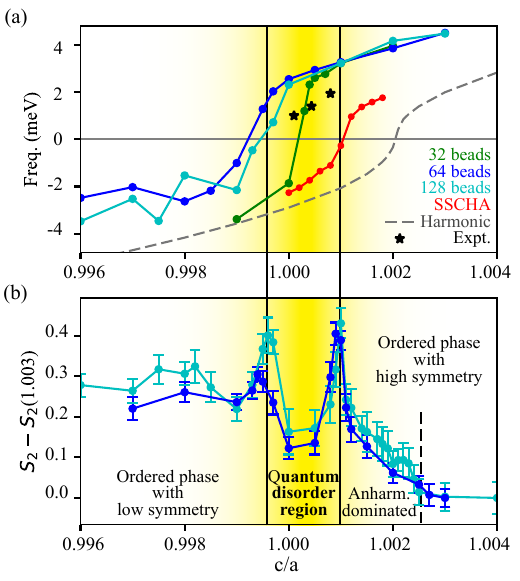}
	\caption{ 
		(a) Gap of AFD mode with varying $c/a$. A structural phase transition is expected when the gap closes. 
		The PIMD approaches convergence with 128 beads and gives distinct critical point compared with SSCHA and harmonic phonon. 
		(b) the R\'enyi entropy by PIMD. The two peaks correspond to two quantum phase transitions (black vertical lines), both important in pinpointing the QD phase, 
which is highlighted by yellow blocks. 
	}
	\label{Fig4}
\end{figure}

By varying the $c/a$ ratio, we found that the gap approaches zero at $c/a=0.9995$ (0.9993) when 128 (64) beads are used in the PIMD simulations at $T=6~\textup{K}$ (Fig.~\ref{Fig4}(a)). 
The closure of this gap means a phase transition from an ordered phase with a statistically low symmetry to a QD phase whose equilibrium structure has a high symmetry. 
This can be confirmed by further calculations of the R\'enyi entanglement entropy $S_\alpha=\frac{1}{1-\alpha}\ln\textup{Tr}\rho_{\textup{A}}^\alpha$
with $\alpha=2$ by PIMD. 
Here, we choose the 12 oxygen atoms in one of the unit cells as the subregion $\textup{B}$ and the rest of the supercell as subregion $\textup{A}$. 
$\rho_{\textup{A}}$ is the reduced density matrix of $\textup{A}$. 
We use the ``replica exchange'' technique in the path-integral\cite{Hastings2010} of $\textup{Tr}\rho_{\textup{A}}^2$ and apply a thermodynamic integration (TI) 
to achieve high-quality sampling.
We also employ a path regularization for TI following the work by Miha Srdinsek\cite{Miha2023} and the Gauss-Legendre quadrature for integration. 
As in Fig.~\ref{Fig4}(b), there is a peak of the R\'enyi entropy $S_2$ at $c/a=0.9996$ for 128 beads and $c/a=0.9994$ for 64 beads. 
The peak represents a divergence of the entanglement entropy and indicates a quantum phase transition. 
The critical points by R\'enyi entropy and by excitation spectra agree well with each other, pinpointing a clear phase transition.
Besides this, after the asymmetric to symmetric phase transition, we note that there is a second peak at $c/a=1.001$. 
This peak marks a quantum phase transition in the high-symmetry region, which has not been proposed as we know in literature. 
Intuitively, when the barrier completely disappears at $c/a=1.002$ (given by the harmonic phonon), the system must have left the disordered 
region because the local degeneracy and QF by tunneling have vanished. 
Therefore, there should be a transition from QD back to ordered phase due to vanishing of local degeneracy. 
However, strangely, this transition happens before the barrier disappears. 
Considering the fact that anharmonicity can lift the frequencies of soft phonon compared with the harmonic results, we also calculate the gap using SSCHA with a $3\times3\times3$ supercell of primitive cell. 
The results are shown by the red line in Fig.~\ref{Fig4}(a).
The gap closes around $c/a=1.001$, which is in excellent consistence with our second peak of $S_2$. 
Therefore, it is the anharmonic effects which bring the transition from $c/a=1.002$ (harmonic phonon) to $1.001$ (SSCHA and our second peak), which accelerates the vanishing 
of QF by tunneling. 
As boundary of tunneling and anharmonic effect, we note that the second transition is important in determining the existence of QD region. 
Our methods on excitation spectra and R\'enyi entropy are able to solve the earlier controversies on QD and anharmonicity, and they are predictive for real materials. 
We now suggest some criteria to quickly filter out probable soft modes possessing QD behavior. 
The first one is a temperate barrier which is necessary for strong QF. 
In some cases, the barrier size and tunneling strength is adjustable by pressure, strain, isotope or doping. 
This may be responsible for some intricate phenomenon in high-pressure or doped systems, such as Ice VII-VIII-X phase transition and superconducting dome in doped $\textup{SrTiO}_3$\cite{benoit1998,Reiter2002,Edge2015}. 
The second one is a weak intersite interaction. 
This may be important for systems composed by heavy atoms such as many perovskites and the ``electron-crystal phonon-glass'' materials\cite{Takabatake2014}. 
There are also hints from experiments, such as a wide nuclear density distribution and optical mode with ultra-low frequency in scattering spectra\cite{Hermann2006,Liu2016}. 
The former has been considered as indication of nuclear tunneling. 
The frequencies, around 1 meV, conform to the energy scale of nuclear tunneling. 
An example is the 0.5 meV inelastic peak in neutron scattering spectra of $\textup{BaTiS}_3$ which is proposed to be related to tunneling and responsible for the anomalous glass-like thermal conductivity\cite{sun2020}. 
In our case, the frequency of AFD mode is around 2 meV and it is close to the value from Raman and neutron spectra\cite{Shirane1969,Cao2000} (black stars in Fig.~\ref{Fig4}(a)). 
Besides these, a feature of special excitation like in quantum spin liquids is also desirable\cite{han2012}.
Ideally, $\textup{SrTiO}_3$ could be approximated to a 3-D quantum Potts or Ising model. 
Although a complete description of the latent quasi-particle is still hard because both theoretical and experimental attempts on 3-D materials are not yet sprouted, we take a step forward to unravel the quantum man-body physics. 
Our method on excitation spectra may give some clues. 
Beyond the adiabatic picture, we also expect that non-trivial coupling with electron and acoustic phonon which may be related to unconventional superconductivity and glass-like thermal conductivity, and we hope the progresses made here can stimulate such investigations. 

\begin{acknowledgments}
	
We are supported by the National Science Foundation of China under Grants No.~12234001, No.~12404257, No.~12474215, and No.~62321004, 
the National Basic Research Program of China under Grants No.~2021YFA1400500 and No.~2022YFA1403500. 
The computational resources are provided by the supercomputer center at Peking University, China.

\end{acknowledgments}

\bibliography{main}

\end{document}